# AVICENNA'S LEGACY IN ASTRONOMY

Rizoi Bakhromzod[1,2*]
[1] S. U. Umarov Physical-Technical Institute, National Academy of Sciences of Tajikistan, 299/1 Ayni Street, Dushanbe 734063, Republic of Tajikistan
[2] Institute of Astrophysics, National Academy of Sciences of Tajikistan
299/5 Aini Street, Dushanbe 734063, Republic of Tajikistan
*e-mail: rizo@physics.msu.ru

**Abstract.** The paper reassesses the largely neglected contribution of Avicenna (Ibn Sina; 980–1037) to medieval tajik-persian astronomy. Drawing on published primary and secondary sources, it reconstructs the main directions of his scientific activity: the construction of an observatory at Isfahan; the design of a high-precision angular instrument that anticipates the modern vernier principle; the formulation of an original method for determining terrestrial longitude from lunar culmination; a systematic refutation of predictive astrology; an optical explanation for the daytime invisibility of the fixed stars; and the earliest extant descriptions of both the transit of Venus on 24 May 1032 and the supernova SN 1006. These achievements not only anticipated comparable European advances by several centuries but also shaped subsequent developments within the Islamic and Latin astronomical traditions. The paper further notes Avicenna's modern scientific commemoration in the naming of asteroid (2755) Avicenna and the lunar crater Avicenna.

## 1 INTRODUCTION

Abu Ali Ibn Sina (980–1037), better known in the West as Avicenna, ranks among the foremost polymaths of human civilisation. Public perception, however, remains one-sided: he is commonly remembered as a physician—and, to a lesser extent, as a poet-philosopher—whereas in reality he was also an accomplished chemist, a brilliant astronomer, an innovative biologist and, more broadly, an outstanding natural philosopher. It is therefore essential to reassess and disseminate the lesser-known scientific facets of his legacy, in particular his astronomical work, which constitutes the principal focus of the present study [Kanefsky 1952; Kloch-kova et al. 2018; Andreev & Ostroushko 2020; Hajar 2013].

Ibn Sina's physical-mathematical and natural-scientific concepts are set out not only in his encyclopaedic summae—Dāneshnāme-ye ʿAlāʾī ("Book of Knowledge"), Kitāb al-Shifāʾ ("Book of Healing"), al-Qānūn fī l-Ṭibb ("Canon of Medicine") and Kitāb al-Najāt ("Book of Salvation")—where he systematised the entire corpus of knowledge available in his time, but also in a series of specialised treatises devoted to arithmetic, geometry, optics, astronomical instrumentation and the structure of the universe: Risāla fī l-Hisāb (Treatise on Arithmetic), Risāla fī l-Zawiyya (Treatise on the Gnomon), Maqāla fī l-Ālāt al-Raṣadiyya (Treatise on Astronomical Instruments), Mukhtaṣar Uqlīdis (Compendium of Euclid), Mukhtaṣar al-Majisṭī (Compendium of the Almagest), Risāla dar Handasa (Treatise on Geometry), Risālat al-Ajsām al-Samāwiyya (Treatise on the Celestial Bodies), al-Mukhtaṣar fī ʿIlm al-Hayʾa (Epitome of Astronomy), Kitāb al-Alwān (Book of Colours), Miʿyār al-ʿUqūl (The Touchstone of Intellect), Qurāḍa-ye Ṭabīʿiyyāt (On "Natural Filings"), Risāla Dhikr Asbāb al-Raʿd wa-l-Barq (On the Causes of Thunder and Lightning), and others [Komili & Muslikhid-dīnov 2021].

The sections that follow examine Avicenna's principal astronomical ideas and achievements in detail.

## 2 IBN SINA'S MAJOR ASTRONOMICAL WRITINGS

Medieval biographical sources—most notably Ibn Sina's own *Autobiography*—relate that the young scholar acquired his first systematic knowledge of astronomy from Ptolemy's *Almagest*, which he studied while attending courses in logic and astronomy under Abū ʿAbdallāh al-Nātilī. Thanks to an extraordinary aptitude for independent study, Ibn Sina often mastered the material more thoroughly than his mentor, to the point of explaining difficult passages of the *Almagest* to al-Nātilī himself (Ibn Sina 1980; Janos 2011).

To date, nine astronomical works can be regarded as authentically Avicennian. They fall into four broad categories: (i) epitomes of Ptolemy's *Almagest*; (ii) treatises on instruments and observational practice; (iii) philosophical and cosmological discourses; and (iv) miscellaneous essays.

1. **Taḥrīr al-Majisṭī** (Compendium of the *Almagest*), composed at Jurjān in 1012–1014 and later revised as Part IV of the mathematical section of the *Shifāʾ*. Two items commonly treated as independent are integral components of this work:
    - *Ibtidāʾ al-maqāla al-muḍāfa ...* ("Beginning of the Treatise Appended to the *Almagest* Epitome Containing Matters Not Explicit in the *Almagest*"), wherein Ibn Sina states that "it behoves us to reconcile what is stated in the *Almagest* with what is known from Natural Science." The appendix discusses the dynamics of celestial motion, reassesses Ptolemy's values for precession and obliquity (Avicenna's own determination is 23° 33′ 30″), proposes an explanation for the motion of the solar apogee, and analyses the latitudinal effects of epicycle poles.
    - *Fī an laysa li-l-arḍ ḥarakat intiqāl* ("That the Earth Possesses No Local Motion"), which reviews Ptolemy's arguments against terrestrial rotation and finds them inadequate.
2. **Al-Arṣād al-kulliyya** (Comprehensive Observations), written at Jurjān (1012–1014) for Abū Muḥammad al-Shīrāzī and incorporated posthumously by Jūzjānī into the *Najāt*. The nine-chapter tract was later translated into Persian as *Raṣadhā-yi kullī* in the *Dānešnāme-ye ʿAlāʾī*.
3. **Maqāla fī l-ālāt al-raṣadiyya** (Treatise on Astronomical Instruments), drafted in Isfahān between 1024 and 1037 during Avicenna's observational programme for ʿAlāʾ al-Dawla.
4. **Fī ṭūl Jurjān** (On the Longitude of Jurjān), dedicated to Zarrayn Kīs, daughter of Shams al-Maʿālī, and written in Jurjān (1012–1014). The text is lost but criticised by al-Bīrūnī in *Taḥdīd al-amākin* for its practical shortcomings.
5. **Al-Samāʾ wa-l-ʿālam** (*De caelo et mundo*), composed for Abū l-Ḥusayn Aḥmad al-Sahlī and probably identical with the chapter of that name in the *Shifāʾ*.
6. **Maqāla fī l-ajrām al-samāwiyya (al-ʿulwiyya)** (Treatise on the Celestial Bodies), likewise cosmological-natural-philosophical rather than mathematically astronomical.
7. **ʿIllat qiyām al-arḍ fī ḥayyizihā (fī wasaṭ al-samāʾ)** (On the Cause of the Earth's Remaining in Its Position at the Centre of the Heavens), written in Gurganj ca. 1005–1012 and also dedicated to al-Sahlī.
8. **Maqāla fī ibṭāl ʿilm [aḥkām] al-nujūm / Risāla fī l-radd ʿalā l-munajjimīn** (Essay on the Refutation of Astrology), which polemicises against predictive astrology and, together with Avicenna’s classification of the sciences, demarcates astronomy from astrological practice.

9. **Maqāla fī khawāṣṣ khaṭṭ al-istiwāʾ** (Treatise on the Characteristics of the Equator), now lost; Avicenna’s view that the equatorial regions are the most temperate is preserved in the *Canon of Medicine* and was debated by al-Bīrūnī, Fakhr al-Dīn al-Rāzī and Naṣīr al-Dīn al-Ṭūsī (Ragep 2007).

These texts collectively attest to Ibn Sina’s sustained engagement with theoretical, observational and instrumental astronomy and form the basis for the analyses presented in the following sections.

## 3 CONSTRUCTION OF THE ISFAHAN OBSERVATORY

At the court of the Buyid amīr ʿAlāʾ al-Dawla Muḥammad (r. 1008–1041) a critical discussion arose concerning the inadequacies of the civil calendar then in use, which still relied on antiquated observations. ʿAlāʾ al-Dawla commissioned Ibn Sina to undertake a new programme of stellar observations and supplied the requisite resources. Together with his pupil Abū ʿUbayd al-Jūzjānī the philosopher established an observatory in Isfahan. A renewed study of Ptolemy’s Almagest convinced them that the extant ephemerides were riddled with errors and that a completely new zij was required. The project pursued two principal aims:

(i) the most accurate possible determination of stellar positions;

(ii) the revision of astronomical tables, especially the ephemerides.

Ibn Sina held that such corrections demanded an independent observational baseline. According to Jūzjānī, construction of the facility took eight years, during which time Ibn Sina devised several innovative instruments and devoted a separate treatise to their description. Owing to the master’s numerous obligations, frequent travels and other practical obstacles, systematic work at the observatory eventually came to an end.

## 4 IBN SINA’S ORIGINAL ANGULAR INSTRUMENT

The *Maqāla fī t-ṭarīq al-ladhī ʿasara*-hu *ʿalā sāʾir al-ṭuruq fī ittikhādh al-ʿalāt al-raṣadīyya* (“Treatise on a Method Preferable to All Others for Constructing Observational Instruments”; Leiden, Univ. Bibl., MS 184/4) outlines a novel device intended to deliver reading accuracy down to one tercia (1/60 arc-second), far surpassing that of the traditional ring, quadrant and armillary sphere (Ahmadov & Vakhabov 1980).

The instrument consists of two hinged limbs, each bounded by four perfectly polished parallel planes.

- The lower, or *heavy*, limb is the principal element; its length is at least seven cubits (~ 3.5 m).
- The upper, *light*, limb is shorter and carries an auxiliary measuring assembly.

On the inner faces of the limbs two converging scales, **LK** and **LS**, are engraved in uniform degree-minute divisions. Sub-minute fractions are obtained by an ingenious attachment:

- a fixed pin **WG** on the light limb supports two parallel sighting plates **H** and **F**;
- a sliding pin **OP** carries plates **Z** and **Q**;
- during observation a grooved spacer **MN** is inserted between the limbs, forming a mobile baseline.

Together these parts realise what is now called the *vernier principle*: the observer first notes the integral division on the main scale and then derives the fractional part from the relative displacement of the auxiliary plates.

The instrument is mounted on a massive cylindrical base whose diameter equals twice the length of the heavy limb. The common hinge rests on an axle at the centre of the cylinder, while the distal end of the heavy limb glides around the 360° rim, graduated down to terciae. After the meridian line is established on a level surface, rotation about the axle yields the azimuth *A*. Raising the light limb and inserting **MN**, the observer sights a celestial object through the H–F–Z–Q dioptrics and determines not the altitude *h* directly but its sine, sinh=MN/LM. If the reading falls between divisions, a supplementary measurement with the upper or lower dioptrics provides a correction, allowing *h* to be computed with sub-minute precision.

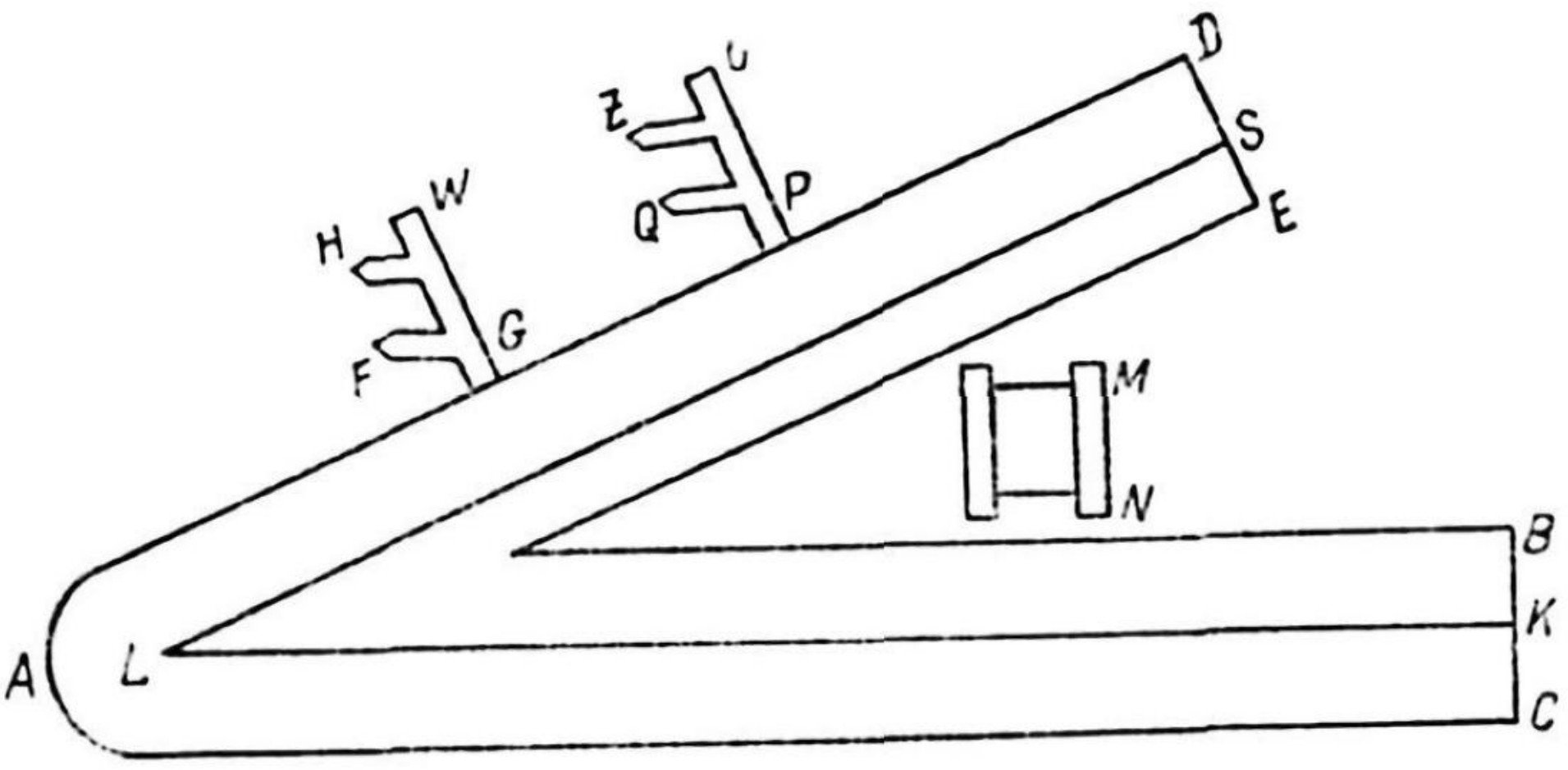

**Fig. 1.** Schematic diagram of Ibn Sina's astronomical instrument.

Ibn Sina thus introduced the vernier method more than five centuries before its independent rediscovery by Pedro Nunes (1542). For equal physical size his "compass" achieved markedly higher accuracy than classical instruments without recourse to unwieldy enlargement. It enabled the determination of stellar altitudes and azimuths with errors below one arc-minute and furnished the practical basis for his lunar-culmination method of finding terrestrial longitude. Avicenna's design therefore represents a significant milestone in the evolution of medieval observational technology (Rosenfeld 1984; Dvoryaninov 2021).

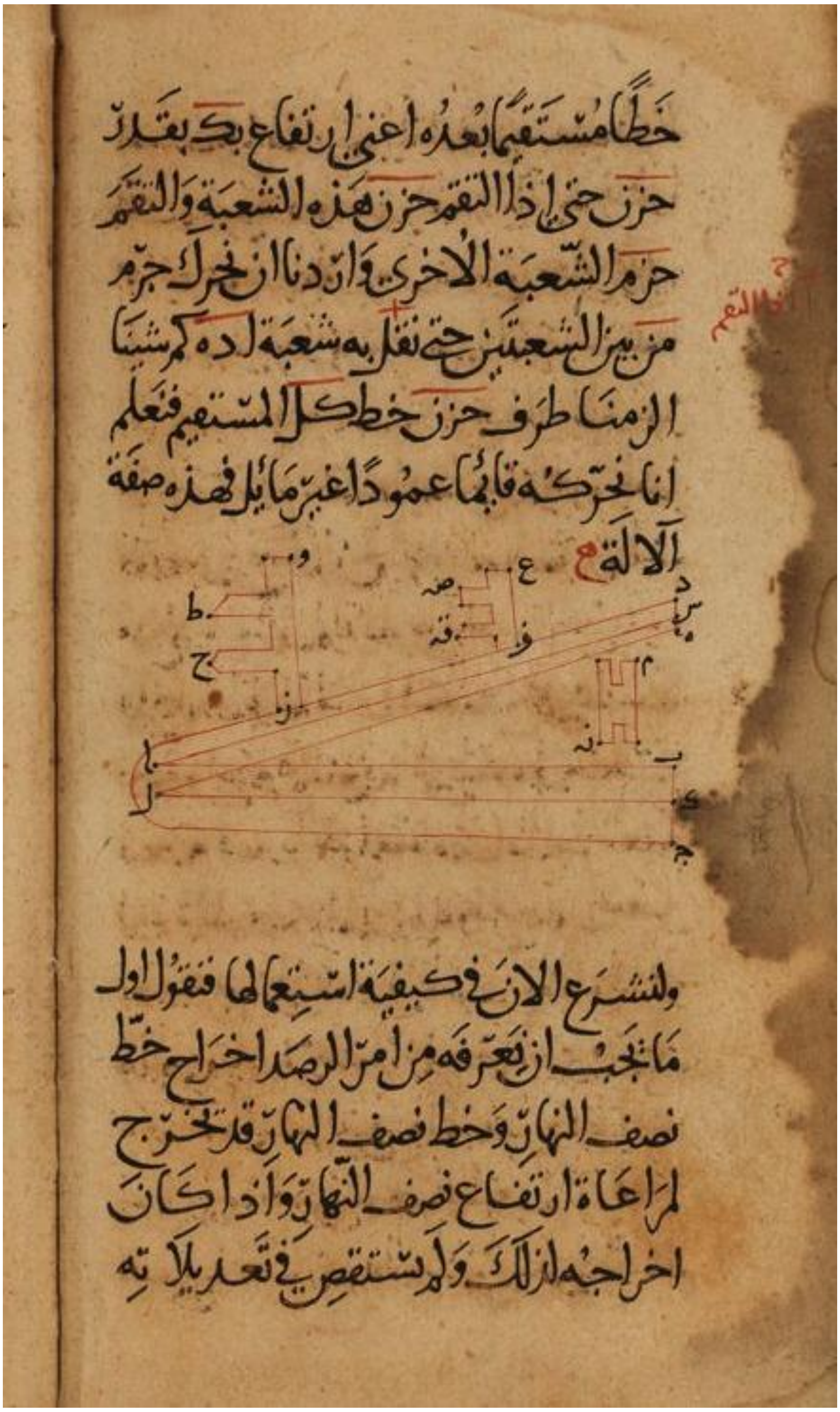

**Fig. 2**. Folio from Ibn Sina’s manuscript, Leiden University Library

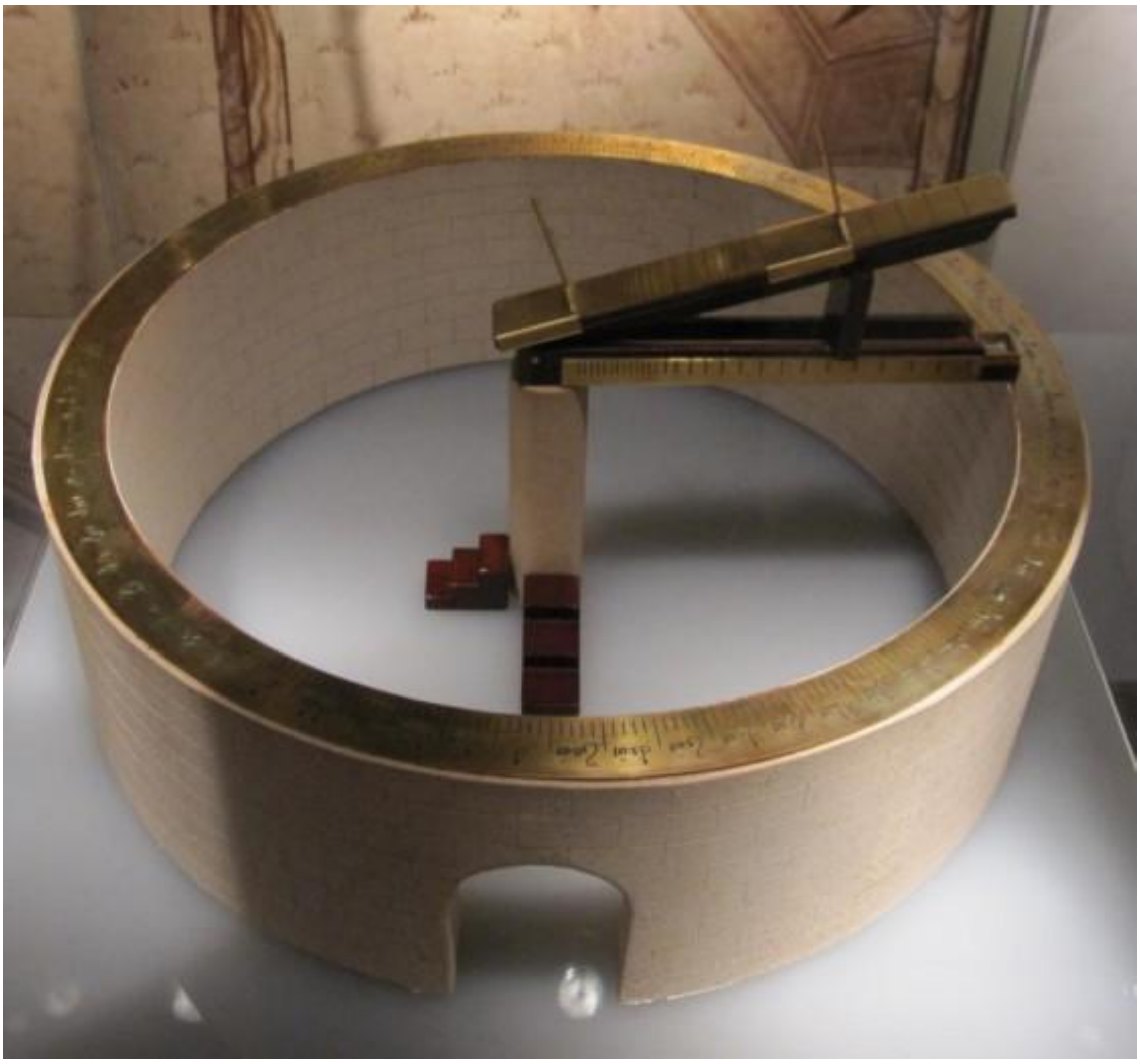

**Fig. 3.** Model of Ibn Sina's astronomical instrument on display at the Istanbul Museum of the History of Science and Technology in Islam

## 5 A NEW METHOD FOR DETERMINING GEOGRAPHICAL LONGITUDE

Reliable evidence for Ibn Sina's observational activity is provided by his contemporary al-Bīrūnī. In the *Tahdīd al-amākin* ("Geodesy"; Bīrūnī 1966) the latter refers to a now-lost *Epistle to Zarrayn Gīs*, daughter of Shams al-Maʿālī, in which Ibn Sina determines the longitude of the Gurgān capital. The historical setting is clear: in 1012 the scholar moved to Gurgān intending to enter the service of Qābūs ibn Wušmagīr. After Qābūs's assassination power passed to his son Manūchihr, while considerable influence was wielded by Manūchihr's sister Zarrayn Gīs, who commissioned Ibn Sina to refine the city's longitude. The absence of a lunar eclipse that year and the impossibility of arranging simultaneous observations with an external station of known longitude compelled him to devise an entirely new procedure.

The algorithm was as follows. — (i) The meridian altitude of the Moon was measured in Gurgān. (ii) Using astronomical tables computed for Baghdad, the expected altitude was calculated for a provisional longitude difference (initially 8°). (iii) Observed and computed altitudes were compared; by iteratively adjusting the assumed difference and recomputing, Ibn Sina obtained agreement when an additional 1° 20′ was added. The final Baghdad–Gurgān offset thus amounted to 9° 20′. Al-Bīrūnī emphatically assigns sole authorship of the method to Ibn Sina, underscoring its innovative character for the period's demanding practical astronomy (Bulgakov 1980).

The lunar-culmination technique, introduced shortly after 1012, anticipated European developments by five centuries: only in 1514 did Nicolaus Werner independently propose the same procedure in his commentary on the *Almagest*. Al-Bīrūnī's testimony therefore augments Ibn Sina's image, revealing not merely a philosopher-physician but also a field astronomer who contributed a novel solution to one of medieval geodesy's most difficult problems.

## 6 WHY THE STARS ARE INVISIBLE BY DAY: AVICENNA'S EXPLANATION

Avicenna rejected the Aristotelian notion that the stars shine only by reflecting sunlight and asserted that they are self-luminous. (He initially extended this view to the planets, a point refuted by modern observations; yet his judgment concerning stellar light proved correct.)

The immediate stimulus for a separate treatise on the subject was a question posed by the ruler Ghiyāth al-Dīn: "Why do we not see the stars during the daytime?" Dissatisfied with prevailing answers, Avicenna composed a concise essay that has survived in manuscript form. He begins by noting that the air filling all space around us is transparent; a transparent medium neither emits light nor becomes visible while transmitting it.

He therefore dismisses the idea that the sky itself glows by day and darkens at night. Sunlight, he argues, traverses the atmosphere unhindered, and in daylight the brilliance of the Sun, together with light diffusely reflected from terrestrial objects, overwhelms human vision, depriving us of the ability to discern the faint stellar beams. If the atmosphere itself emitted light, distant objects on Earth would be obscured: in a desert or mountain landscape one could not distinguish a far-off cliff or wall.

Avicenna attributes the scattering of sunlight to impurities—water vapour and dust—suspended in the air; the cleaner the atmosphere, the easier it is to observe the stars. To illustrate that intense illumination, rather than any intrinsic property of the heavens, hampers stellar visibility he offers the example of a bonfire: when standing close to the flames the stars vanish from sight, but as one moves away they reappear (Ünver 1946).

Thus, centuries before the modern theory of atmospheric optics, Avicenna explained the diurnal invisibility of the stars as a combined effect of solar brightness and aerial scattering.

## 7 WHY ASTROLOGY IS NOT A SCIENCE: IBN SINA'S EPISTEMIC CRITERIA

In the medieval world astrology—ṣanʿat al-tanjīm, the "star-craft"—enjoyed unrivalled prestige: the positions of Sun, Moon, planets and fixed stars were consulted to wage wars, contract marriages and found cities, and almost every Eastern or European ruler retained a court astrologer. Against this backdrop Ibn Sina's critique marked a turning-point in the history of scientific scepticism, for he was the first to offer a systematic, logic- and evidence-based refutation of predictive astrology. His principal anti-astrological work, Ishāra bar fasād ʿilm aḥkām al-nujūm ("Indication of the Unsoundness of the Science of Stellar Judgements"), though brief, is densely argued, treating physical principles, epistemic limitations and the ethical dangers of blind faith in horoscopes.

### 7.1 Rejection of the theoretical foundations

Ibn Sina dismisses the basic astrological premise that planets or zodiacal signs possess intrinsically "good" or "evil" natures. The heavens, he insists, constitute an integrated, harmonious organism; to assign moral qualities to particular sectors is as absurd as declaring each breath of wind virtuous or vicious (Harvey 2019). Causal claims must be measurable: if an influence cannot be quantified, it cannot qualify as scientific fact.

### 7.2 Limits of human cognition

While not excluding a priori that cosmic processes may affect the sublunary world—including climate cycles—Ibn Sina stipulates three conditions for demonstrating such influence: (i) regular observation, (ii) quantitative correlation and (iii) elimination of confounding factors. Astrology, he argues, meets none of these requirements: the stellar realm is too remote and terrestrial variables too complex to yield unequivocal formulas of destiny.

### 7.3 Epistemology: reliable knowledge versus fiction

The philosopher distinguishes three strata of "knowing":

(a) Rational-empirical (burhānī)—grounded in logic, observation and experiment;

(b) Intuitive—acquired through prophetic revelation, reliable yet context-bound for Muslims;

(c) Illusory—including astrology, magic and divination, which trade on statistical coincidence and psychological suggestibility, mistaking subjective correlations for causation (Black 2013).

Deliberately opposing rigorous mathematical astronomy to speculative astrology, Ibn Sina cites Ptolemy's Almagest as a benchmark of objective knowledge, showing that true astronomy can predict eclipses to the minute, whereas astrology fails even retrospectively to account for many historical disasters. His logical framework was later adopted by Moses Maimonides in the Tractatus de astrologia and, via Latin translations by Gerard of Cremona and Michael Scot, influenced Thomas Aquinas and Albertus Magnus. Within Islam the critique was extended by Naṣīr al-Dīn Ṭūsī and Quṭb al-Dīn Shīrāzī, who highlighted the mathematical contradictions in horoscopy.

For Ibn Sina astrology was not merely theoretically unsound but socially harmful: belief in predetermined fate diminishes personal responsibility and facilitates manipulation. He urged rulers to rely on observation, statistics and the counsel of natural philosophers, not on fortune-tellers. His critique stands as one of the earliest systematic manifestations of scientific scepticism toward "star-divination," shaping Muslim, Jewish and Christian intellectual traditions alike and drawing a methodological line between verifiable knowledge and speculation. Avicenna's legacy thus remains a paradigm of rational engagement with popular yet unscientific doctrines.

## 8 CORRESPONDENCE AND DEBATE WITH ABŪ RAYḤĀN AL-BĪRŪNĪ

A pivotal source for reconstructing Ibn Sina's astronomical outlook is his epistolary exchange with the other towering polymath of the age, Abū Rayḥān al-Bīrūnī. According to internal evidence and later scholarship, the dialogue took place when Ibn Sina was about eighteen years old and al-Bīrūnī about twenty-five. The surviving letters (Bīrūnī & Ibn Sina 1973) display an extraordinary level of intellectual acuity: within them the correspondents tackle pressing issues of philosophy, physics and astronomy.

The dossier comprises ten questions on Aristotle's *De Caelo* and eight on his *Physics*. Writing from a broadly Democritean stance, al-Bīrūnī challenges Ibn Sina with a sequence of objections; the latter, adopting an Aristotelian perspective, replies with rigorously argued rejoinders. Despite their youth, the sophistication of the debate attests to the depth of their learning. Both interlocutors address one another with respect, yet defend their positions in uncompromising logical terms.

Themes range from the nature of the celestial spheres and the dynamics of motion to the structure of the cosmos, the theory of light, the physiology of vision and transparency, the divisibility of matter and the possible plurality of worlds (Hullmeine 2019). The correspondence mirrors the intellectual climate of the early eleventh century, revealing how thoroughly Aristotelian ideas were assimilated and critically examined. It likewise illustrates the methodological unity of logic, mathematics and natural philosophy, and exemplifies the "democratic" character of medieval scientific discourse—dialogue grounded in mutual critique and reasoned argument.

## 9 IBN SINA'S OBSERVATION OF THE VENUS TRANSIT

One of Avicenna's most remarkable astronomical achievements was the first recorded observation of a transit of Venus across the solar disc. Naṣīr al-Dīn al-Ṭūsī reports that in his *Taḥrīr al-Majisṭī* the philosopher noted seeing Venus on the face of the Sun "as a dark spot", an event that occurred on 24 May 1032 (Julian) and recurs only about a dozen times per millennium. Europe did not witness so rare a phenomenon until the seventeenth century.

Scepticism arose in the twentieth century when Goldstein (1969), basing himself on Jean Meeus's 1958 calculations, suggested that Ibn Sina might have mistaken a solar spot for the planet. Modern ephemerides and numerical modelling, however, demonstrate that from Hamadān or Isfahān the transit was perfectly visible to the unaided eye near sunset, when solar glare is attenuated; Kapoor (2013) has therefore vindicated the historicity of the medieval report.

The observation carried profound cosmological implications. In the Platonic–Aristotelian geocentric scheme many scholars still located the Venusian sphere outside the solar one. Ibn Sina's testimony showed that Venus lies between Earth and Sun and is a finite material body—an inference that paved the way for the later Copernican heliocentric revolution (Özel & Budding 2024).

Furthermore, in the *Kitāb al-Shifāʾ* he explained the rarity of such events by noting that the orbit of Venus is inclined to the ecliptic at a fixed angle and intersects it only at its nodes—effectively recognising the concept of orbital nodes centuries before it became standard in planetary theory.

## 10 IBN SINA AND THE FIRST DESCRIPTION OF A SUPERNOVA

Another astronomical feat attributed to Ibn Sina—recognised only in the past decade—is his record of a stellar outburst that modern scholarship identifies as a supernova. In the *Kitāb al-Shifāʾ* he reports that in 397 AH (AD 1006) an exceptionally bright "new star" appeared in the sky and remained visible for roughly three months. He remarks that its brilliance waned from day to day: at first it glowed with a dim greenish hue, then flared whiter than usual, and finally faded from sight.

For many years historians assumed that the account referred to a comet. In 2016, however, three German researchers showed that Ibn Sina had witnessed the supernova **SN 1006** in the constellation Lupus, whose peak brightness occurred in May 1006. The remnant of this event—radio source PKS 1459-41—was not discovered until 1965; it lies ≈ 7100 ly from Earth and is still expanding, with an angular diameter of ~ 30′ (comparable to the full Moon) and a physical diameter of ≈ 60 ly (Neuhaeuser et al. 2017).

SN 1006 is the brightest supernova ever recorded by humankind: as a Type Ia event it outshone Venus by a factor of ~ 16 and appeared 2.5–3 times larger in angular size; for several weeks it was even visible in daylight. Besides Ibn Sina, the outburst was observed by ʿAlī ibn Riḍwān in Egypt, Zhōu Kēmín in China, and watchers in Yemen and Japan (Winkler et al. 2003; Stephenson 2010).

Ibn Sina's chronicle thus stands as the earliest surviving description of a supernova in Islamic literature, underscoring his role as a careful sky-watcher and enriching the global historical record of transient celestial phenomena.

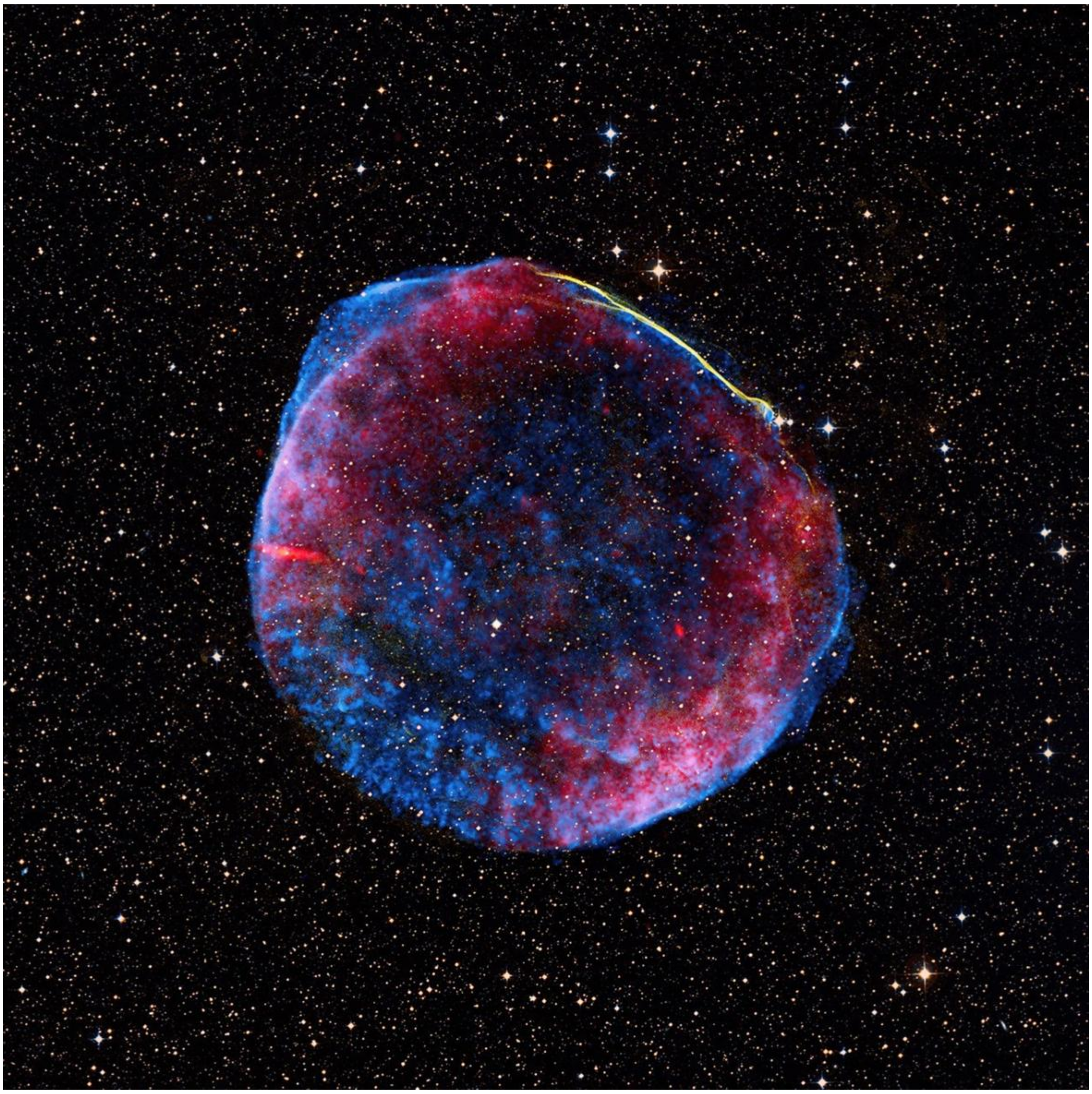

Fig. 4. Chandra X-ray image of SN 1006, highlighting multimillion-degree gas in red/green and high-energy electrons in blue from the white-dwarf supernova first seen as a “new star” in 1006 CE.

## 11 MODERN COMMEMORATION IN CELESTIAL NOMENCLATURE

In recognition of Ibn Sina’s enduring scientific legacy, his name today adorns not only cities, districts, universities and streets, but also two officially designated celestial bodies.

**Asteroid (2755) Avicenna** (provisional designations 1973 $SJ_4$, 1978 $UX_1$) was discovered on 26 September 1973 by the Soviet astronomer Lyudmila I. Chernykh at the Crimean Astrophysical Observatory and formally named for Ibn Sina on 28 March 1983. Its perihelion distance is 316.769 million km and it completes one revolution about the Sun in 4.8 years; the estimated diameter is ~ 11 km (Minor Planet Center database).

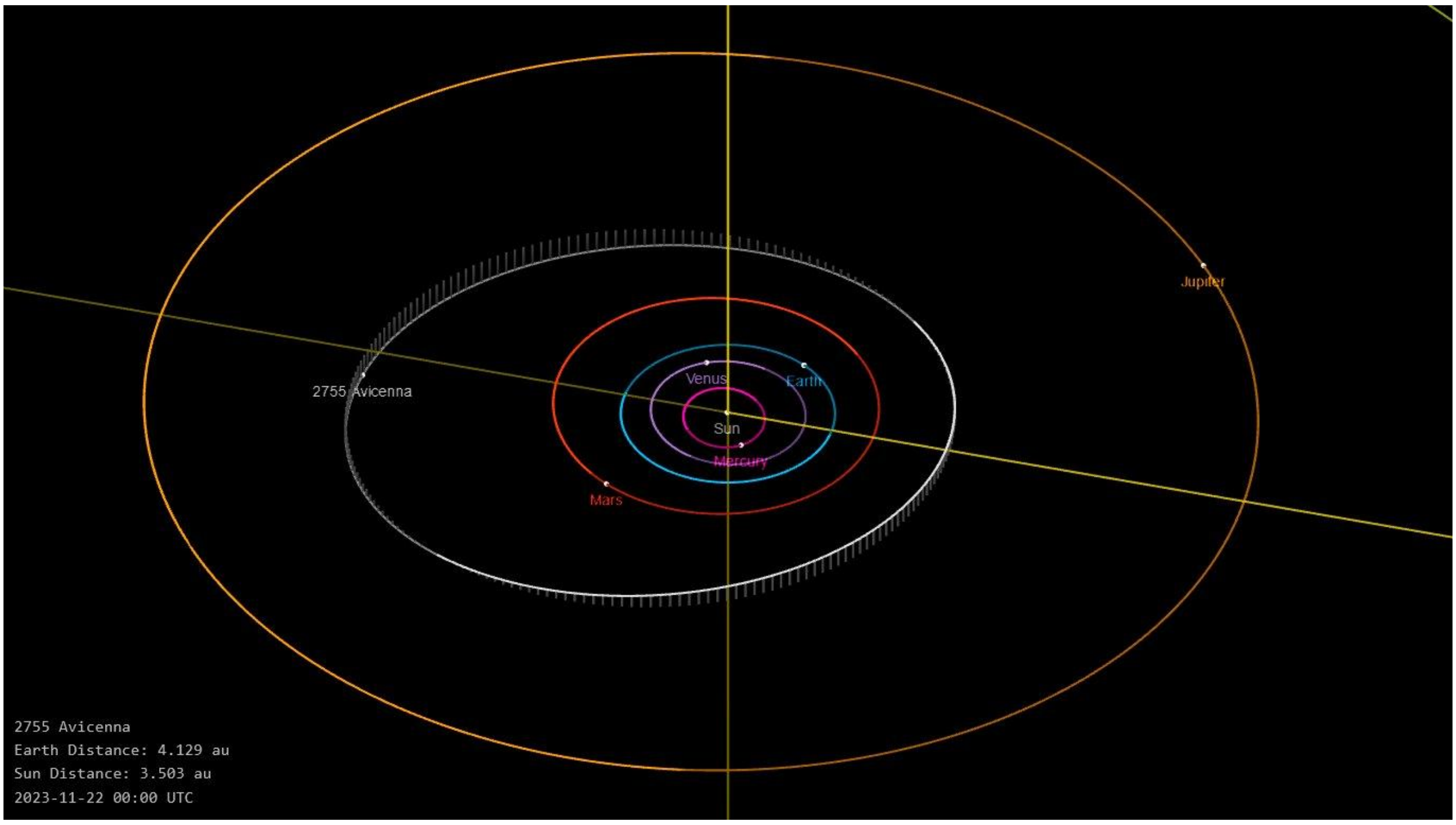

**Fig. 5.** Orbital diagram of asteroid 2755 Avicenna from the JPL Small-Body Database Viewer (VOPDA view).

**Avicenna crater**, situated on the lunar far side at 39° 38′ N, 97° 17′ W, measures 73 km in diameter and ~ 2.7 km in depth. The International Astronomical Union (IAU) adopted the eponym in 1970 as part of its systematic lunar nomenclature (Menzel et al. 1971; Hoenig 2021).

These designations ensure that the memory of the Tajik-Persian polymath is literally inscribed upon the heavens he so assiduously studied.

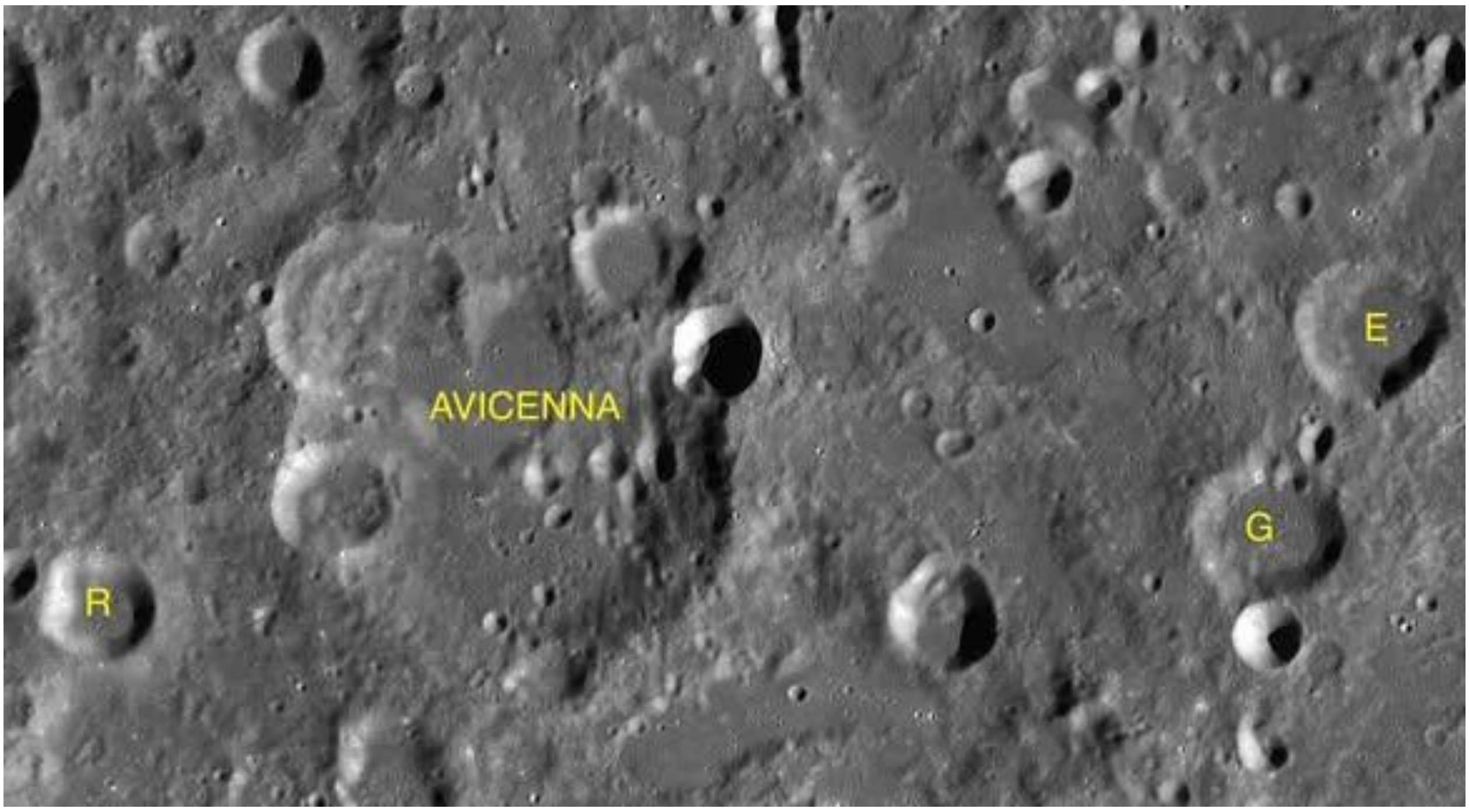

**Fig. 6.** Avicenna sattelite craters map

## 12 CONCLUSIONS

The foregoing analysis shows that Avicenna not only devised an angular instrument employing a vernier-type scale—achieving record observational accuracy for his day—but also introduced an original method for determining terrestrial longitude from lunar culmination, anticipating comparable European solutions by five centuries. His observation of the 24 May 1032 transit of Venus provided empirical evidence that the planet moves between Earth and Sun, while his chronicle of the supernova SN 1006 constitutes the earliest reliable report of such a phenomenon; together these findings undermined traditional geocentric assumptions. The correspondence with al-Bīrūnī displays an early synthesis of logical analysis, experimental reasoning and mathematical computation, establishing a methodological foundation for subsequent Islamic science. Avicenna's systematic refutation of astrology set a precedent for rational opposition to star-divination and influenced Jewish, Christian and later Islamic thought. Finally, the modern attribution of his name to asteroid 2755 Avicenna and the lunar crater Avicenna attests to the enduring significance of his contributions to the history of science.

## 13 ACKNOWLEDGEMENTS

The author gratefully acknowledges the financial support provided by Mr Zafar Mirzo and thanks Dr Sherali Rizoyon (PhD, Political Science) for his valuable suggestions that substantially improved this paper.